\documentclass[a4paper,11pt]{article}
\pdfoutput=1 
\usepackage{jinstpub}
\usepackage{lineno,hyperref}
\modulolinenumbers[5]
\usepackage{graphicx}
\usepackage{subcaption}
\usepackage{amsmath}   
\usepackage{lineno} 
\usepackage[english]{babel}
\usepackage{siunitx}
\usepackage{multirow}
\usepackage[dvipsnames]{xcolor}

\title{Commissioning of the PADME experiment with a positron beam}
\author[a]{P.~Albicocco,}
\author[b]{R.~Assiro,}
\author[a]{F.~Bossi,}
\author[c]{P.~Branchini,}
\author[a]{B.~Buonomo,}
\author[d]{V.~Capirossi,}
\author[a]{E.~Capitolo,}
\author[a]{C.~Capoccia,}
\author[e,b]{A.P.~Caricato,}
\author[a]{S.~Ceravolo,}
\author[b,1]{G.~Chiodini,\note{Corresponding author}}
\author[a]{G.~Corradi,}
\author[a]{R.~De Sangro,}
\author[a]{C.~Di Giulio,}
\author[a]{D.~Domenici,}
\author[f]{F.~Ferrarotto,}
\author[g,f]{S.~Fiore,}
\author[a]{G.~Finocchiaro,}
\author[a]{L.G~Foggetta,}
\author[h]{A.~Frankenthal,}
\author[a]{M.~Garattini,}
\author[i,j]{G.~Georgiev,}
\author[a,2]{F.~Giacchino,\note{presently at INFN sez. Roma 2, via della Ricerca Scientifica 1, Roma}}
\author[a]{A.~Ghigo,}
\author[a]{P.~Gianotti,}
\author[e,b]{F.~Gontad,}
\author[d]{F.~Iazzi,}
\author[j]{S.~Ivanov,}
\author[j]{Sv.~Ivanov,}
\author[j,a]{V.~Kozhuharov,}
\author[f]{E.~Leonardi,}
\author[k]{E.~Long,}
\author[l,a]{M.~Martini,}
\author[e,b]{M.~Martino,}
\author[e]{A.~Miccoli,}
\author[e,b]{I.~Oceano,}
\author[e,b]{F.~Oliva,}
\author[k]{G.C. Organtini,}
\author[d]{F.~Pinna,}
\author[k]{G.~Piperno,}
\author[e,b]{C.~Pinto,}
\author[k]{M.~Raggi,}
\author[f]{F.~Safai Tehrani,}
\author[a]{A.~Saputi,}
\author[a]{I.~Sarra,}
\author[a]{B.~Sciascia,}
\author[j]{R.~Simeonov,}
\author[a]{T.~Spadaro,}
\author[e,b]{S.~Spagnolo,}
\author[a]{E.~Spiriti,}
\author[c]{D.~Tagnani,}
\author[j]{l.~Tsankov,}
\author[a]{C.~Taruggi,}
\author[f]{P.~Valente,}
\author[f]{A.~Variola,}
\author[a]{E.~Vilucchi}

\affiliation[a]{INFN Laboratori Nazionali di Frascati,\\ via E. Fermi 54, Frascati, Italy}
\affiliation[b]{INFN sez. Lecce,\\ via Provinciale per Arnesano, Lecce, Italy}
\affiliation[c]{INFN sez. Roma 3,\\ via della vasca navale 84, Roma, Italy}
\affiliation[d]{DISAT Politecnico di Torino and INFN sez. Torino,\\ C.so Duca degli Abruzzi 24, Torino, Italy}
\affiliation[e]{Dip. Mat. e Fisica Salento Univ.,\\ via Provinciale per Arnesano, Lecce, Italy}

\affiliation[f]{INFN sez. Roma 1,\\ p.le A. Moro 2, Rome, Italy}
\affiliation[g]{ENEA Frascati,\\ via E. Fermi 45, Frascati, Italy}
\affiliation[h]{Physics Dep. Princeton Univ.,\\ Washington Road, Princeton, USA}
\affiliation[i]{INRNE Bulgarian Accademy of Science,\\ 72 Tsarigradsko shosse Blvd., Sofia, Bulgaria }
\affiliation[j]{Sofia Univ. ``St. Kl. Ohridski'',\\ 5 J. Bourchier Blvd., Sofia, Bulgaria}
\affiliation[k]{Dip. di Fisica Sapienza Univ.,\\ p.le A. Moro 2, Roma, Italy}
\affiliation[l]{DSI Marconi Univ.,\\ via Plinio 44, Roma, Italy  }

\emailAdd{gabriele.chiodini@le.infn.it}

\abstract{

The PADME experiment is designed to search for a hypothetical dark photon $A^{\prime}$ produced in positron-electron annihilation using a bunched positron beam at the Beam Test Facility of the INFN Laboratori Nazionali di Frascati. The expected sensitivity to the $A^{\prime}$-photon mixing parameter $\epsilon$ is 10$^{-3}$, for  $A^{\prime}$ mass $\le$ 23.5 MeV/$c^{2}$ after collecting $\sim 10^{13}$ positrons-on-target.

This paper presents the PADME detector status after commissioning in July 2019. 
In addition, the software algorithms employed to reconstruct physics objects, such as photons and 
charged particles, and the calibration procedures adopted are  illustrated in detail. 
The results show that the experimental apparatus reaches the design performance,
and is able to identify and measure standard 
electromagnetic processes, such as positron bremsstrahlung and  electron-positron annihilation into two 
photons.

}

\keywords{Dark Matter detectors (WIMPs, axions, etc.), 
Large detector-systems performance, Solid state detectors, Spectrometers
}
\usepackage{natbib}

\begin{document}

\maketitle
\flushbottom

\section{Introduction\label{sec:Introduction}}

A new weakly-coupled  $U_{D}$(1) hidden symmetry, with its corresponding dark photon, has been proposed by several extensions of the Standard Model as a possible explanation for the difficulty of detecting dark matter \cite{darksector}.
The Positron Annihilation into Dark Matter Experiment (PADME), at the INFN Laboratori Nazionali di Frascati (LNF) in Italy, searches for a possible dark photon $A^{\prime}$ produced in the process $e^{+}e^{-}\rightarrow\gamma A^{\prime}$, where the $A^{\prime}$ goes undetected \cite{padme}.
A positron beam with maximum energy of 550 MeV, produced and accelerated by the LNF LINAC, hits a low-$Z$ target. The annihilation of a positron with an atomic electron in the target can produce an $A^{\prime}$ with  mass up to $\sim$ 23.5 MeV/$c^2$. Under the assumption that the $A^{\prime}$ escapes detection, due to either being long-lived or decaying predominantly via invisible modes, the production signature is a single photon plus missing mass. PADME is the first experiment to search for such a signal using this technique and would also be sensitive to any new particle produced in $e^{+}e^{-}$ annihilation, including axion like particles, dark Higgs, etc. This paper describes the apparatus and tools developed to accomplish this program, and reports the achieved performance during the early commissioning phase.

\section{The BTF positron beams}
\label{sec:beam}
The positron beam for the PADME experiment is provided by the LINAC of the LNF DA$\Phi$NE complex. The LINAC can deliver bunched beams of electrons or positrons with a maximum repetition rate of 50 Hz which can be sent, via a dedicated beam line, to the experimental halls of the Beam Test Facility (BTF). 

For the PADME physics program, the LINAC must be operated in a dedicated mode and with a specific setup of the BTF beam line:  a bunch duration of hundreds of ns, and an angular spread of few mrad.
Two types of positron beam are available at the BTF: a primary positron beam, produced at the LINAC positron converter, or a secondary positron beam produced at a secondary target installed on the BTF line  \cite{Buonomo:2017yml}\cite{Valente:2017mnr}. 

For primary beams, positrons are generated at the LINAC positron converter located downstream of the first accelerating sections and then transported to the experimental halls. The bunch multiplicity can be varied between $10^{4}$ and $10^{10}$ particles/bunch with maximum energy of 550 MeV and energy spread below 1\%. 

For secondary beams, 
a copper target is installed on the BTF transfer line. 
The interaction of the primary electrons with the target 
produces a broad distribution of particles both in energy and angle. 
These are then momentum-selected and collimated, to produce a lower intensity beam, down to a single particle per spill. 
In this configuration, the maximum energy available for the positrons in the BTF halls is higher, since all the LINAC accelerating sections are used, increasing the maximum energy by 220-250 MeV to a total of 770-800 MeV.

To obtain a higher beam energy, thus increasing the explorable $A^{\prime}$ mass range, a secondary positron beam of energy 545 MeV was used in the first PADME runs of October and November 2018. The sensitivity obtained running with this setup was discovered to be limited by the beam background in the experiment. Therefore, only primary positron beam was used in the data-taking runs of 2019 (this paper) and 2020 (runs with improved beam line) \cite{bossi2022padme}.
For regulars runs, the primary beam energy was set at 490 MeV and 432.5 MeV, respectively for the two periods, and not at 550 MeV, maximum energy, to avoid LINAC instability at top energy.

\section{The PADME detector}
\label{sec:det}
\par

\begin{figure}[h]
\centering
\includegraphics[width=1.0\linewidth]{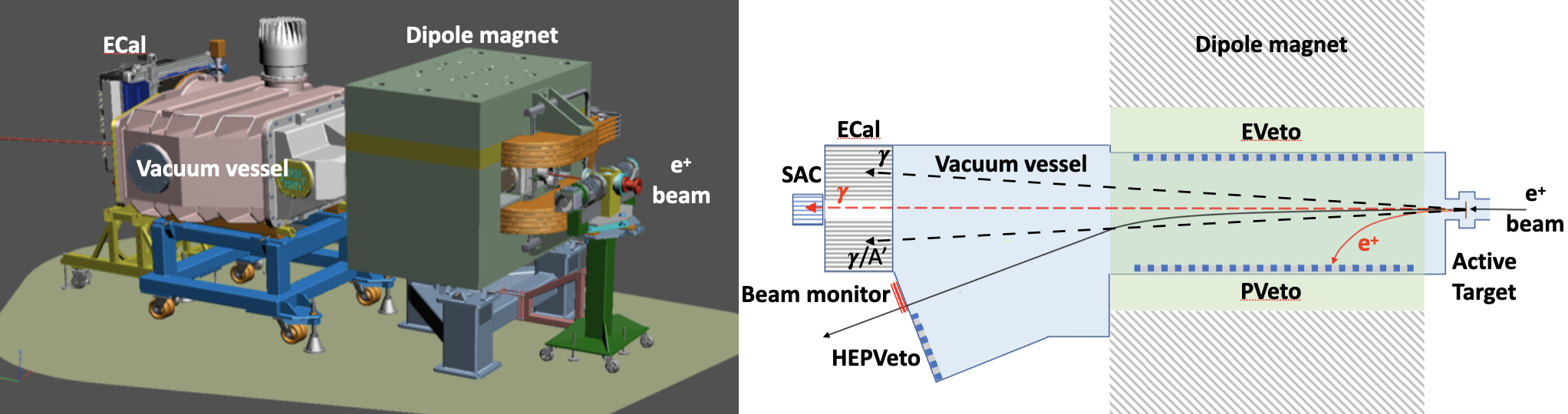}
\caption{
The PADME detector 3D CAD design (Left) and its schematic layout from the top (Right). The positron beam enters from the right. The trajectory of a non-interacting $e^+$ is drawn in black, and in black, the trajectories corresponding to a $\gamma \gamma$/$\gamma A'$ production in the target.  The particle trajectories from a bremsstrahlung interaction in the target are drawn in red.}
\label{fig:padme}
\end{figure}

A sketch of the PADME setup is shown in Fig.~\ref{fig:padme}.
This layout is intended primarily to measure the 4-momentum of the outgoing photons and of the forward-emitted particles after the positrons interact with the target. 
Assuming that the target electrons are at rest, the 
missing mass from interactions with a single photon in the final state can be computed to search for the $A^{\prime}$. 

The main component of the PADME setup is a calorimeter consisting
of two systems: the Electromagnetic Calorimeter (ECal), 
which gives precise measurements of the energy
and impact point of all photons emitted in the angular range 15-82 mrad with respect to the beam direction, and the Small-Angle Calorimeter (SAC) which allows photons emitted below $\sim$15 mrad to be vetoed. The SAC is most sensitive to background reactions such as bremsstrahlung and $e^+e^-\rightarrow 3\gamma$ annihilation. 
\par
Any positron that does not interact in the target is directed towards a beam monitor and the beam dump by a magnetic field, which also deflects secondary charged particles towards a set of dedicated detectors.
These detectors act both as vetoes for background reactions and as spectrometers for forward emitted particles.



\subsection{Active diamond target}
A thin active diamond sensor not only provides the target for the annihilation process, but also measures the XY profile of the beam and the beam multiplicity, {\it i.e.} the number of positrons contained in each bunch.
To have the highest possible ratio of annihilation cross-section to bremsstrahlung cross-section, the lowest possible atomic number Z is required. Out of a range of solid materials considered for the target, diamond (Z~=~6) was eventually chosen because of its relatively straightforward manufacturing process. The sensor has an active surface of 20$\times$20~mm$^2$ and a thickness of 100~$\mu$m. Graphite strips with pitch 1~mm, including a 0.15~mm gap, were created using a laser technique on both diamond faces ~\cite{PaperGraphitization}.
Nineteen strips on one face were oriented horizontally to measure the X view and nineteen strips on the other face were oriented vertically to measure the Y view).
Only 16 strips for each views were connected to the electronic readout~\cite{Oliva:2019alx}.

\subsection{Electromagnetic calorimeter system}


The ECal and the SAC were designed to study different types of interactions.
The main objective of the ECal is to detect photons produced in $e^+e^-$ annihilation. These include photon pairs coming from $e^{+}e^{-}\rightarrow \gamma\gamma$ interactions as well as single photons that may be produced along with the $A^{\prime}$ bosons (see Sec. \ref{sec:Introduction}). 
The ECal is a high resolution detector composed of $616$ scintillating bismuth germanate (BGO) crystals each of dimensions $21\times21\times230\,~\text{mm}^{3}$, arranged in a cylindrical shape and readout by HZC XP1911 photomultipliers. It is placed $\sim$3.4~m downstream from the target and the external radius is approximately $29\,\text{cm}$. The resulting angular coverage ranges from $15.66\,\text{mrad}$ to $82.11\,\text{mrad}$.
The main component of its scintillation light has a decay time of $\sim300\,\text{ns}$. Thus, to avoid pile-up created by the high rate of bremsstrahlung photons which are predominantly emitted in the forward direction, the ECal was built with a hole of dimension 5$\times$5 crystals ($105\times105\,\text{mm}^{2}$).
The energy resolution of the ECal was evaluated to be 2.62\% at $490\,\text{MeV}$~\cite{Albicocco:2020vcy}. 

The main purpose of the SAC is to veto photons from 3$\gamma$ annihilation and bremsstrahlung ($e^{\pm} N \rightarrow e^{\pm} N \gamma$). 
It is placed behind the central hole of the ECal and consists of a matrix of $25$ PbF$_2$ crystals, each measuring $30\times30\times140~\text{mm}^3$ and coupled to Hamamatsu R13478UV photomultipliers. The readout signal from PbF$_2$ crystals comes from Cherenkov light scintillation.
The in-bunch bremsstrahlung rate in this detector was expected to reach $\sim10^{2}\,\text{MHz}$, requiring a double-peak separation better than $5\,\text{ns}$. Typical signals from the SAC are $2-3\,\text{ns}$ long, matching the requirement. 

A beam test of a single SAC unit was performed at the BTF before the detector was constructed. A single-particle electron beam with an energy between $100\,\text{MeV}$ and $400\,\text{MeV}$ was fired onto the PbF$_2$ crystal, to study energy and time resolution, as well as the linearity of the detector response. The measured energy resolution was below 20\% across the entire energy range, and the time resolution was 80 ps. 

\subsection{Charged particle detectors}
Three stations of charged particle detectors are placed inside the vacuum chamber of the experiment \cite{8330040}. They are all composed of bars of polystyrene plastic scintillator measuring $10\times10\times178~\text{mm}^3$, each equipped with a 1.2 mm$^2$ diameter BCF-92 wavelength shifter fibre (WLS fibre) along a longitudinal groove. 
The light from the WLS fibre is detected by a $3\times3$~mm$^2$ Hamamatsu S13360  silicon photomultiplier (SiPM). 
Four SiPMs are mounted on a custom-designed front-end electronics card (FEE-card)
providing a controllable high voltage. 
The card also allows independent temperature and current monitoring for each channel. 
Vacuum feed-through flanges ensure the communication and the signal transmission between the FEE-cards and the custom designed controller modules. 

Two of the charged particle detector stations, the positron veto (PVeto, 90 scintillator bars) and the electron veto (EVeto, 96 scintillator bars), are located inside the dipole magnet and are read out from one side of the plastic scintillators. 
The high energy positron veto (HEPVeto, 16 scintillator bars) is placed downstream towards the positron beam exit window. Here the light output is detected from both ends of each bar. 

\section{PADME data acquisition} 
\subsection{Trigger and readout} 
\label{sec:trg}
The PADME data acquisition system \cite{Leonardi:2017ocd} reads data from a total of 897 channels with an acquisition rate of 50 Hz, defined by the LINAC operation frequency.
All channels are connected to a set of 29 CAEN V1742~\cite{Caen:V1742} switched capacitor ADC boards based on the DRS4 chip
consisting of a 1024 cells switched capacitor array (SCA) with a tunable 12-bit sampling rate.
The relatively slow signal produced by the BGO crystals 
and the target integral signal are sampled at 1 GS/s, while the faster signals from the charged particle vetoes and the small angle calorimeter are sampled at 2.5 GS/s. Two CAEN A3818 PCI Express boards collect data from all ADC boards via 8 optical links, each with a maximum throughput of 80 MB/s.

All ADC boards are interfaced with the main PADME Trigger and Timing System (TTS). 
The TTS consists of a main trigger board and two trigger distribution units.
The trigger board can receive up to six NIM-level external triggers (the most important being the BTF trigger from the LINAC, the cosmic ray trigger for ECal calibration, and the random trigger).

\begin{figure}[htbp]
\centering
\includegraphics[width=1.0\linewidth]{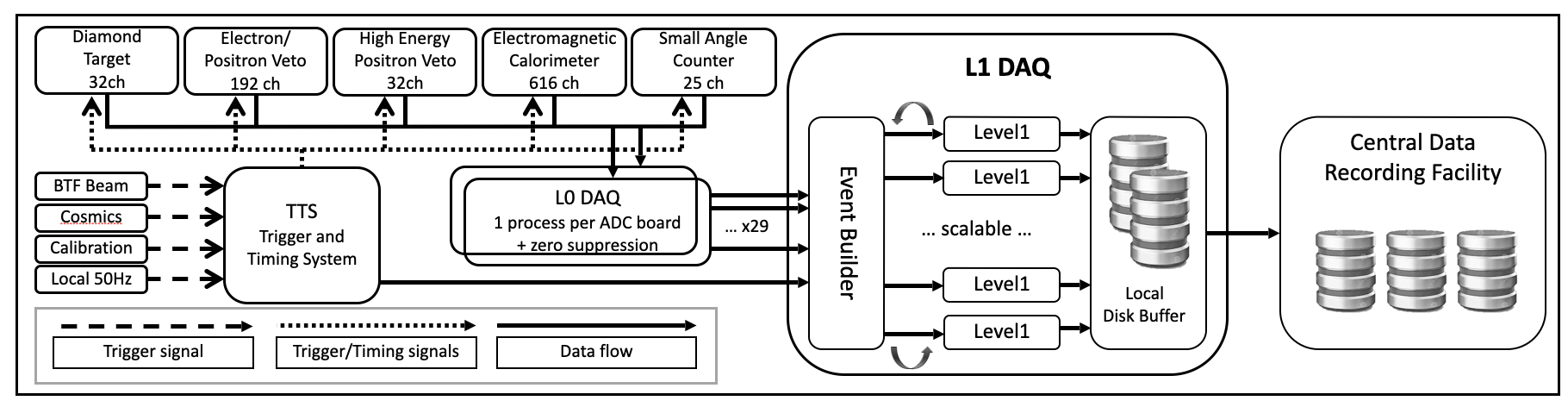}
\caption{Logical scheme of the PADME DAQ system.}
\label{fig:padmedaq}
\end{figure}

The data acquisition process is organized into two stages: level 0 (L0), which collects data from the ADC boards, and level 1 (L1), which merges all data into a single event structure and produces the final data files.
With all sub-detectors enabled, optimal beam conditions, and the zero-suppression algorithm set to flagging mode ({\it i.e.} no data are rejected) the maximum data rate is $\sim$ 50 MB/s.
All data produced by the PADME DAQ system are written to a local disk buffer directly connected to the L1 servers. The total size of the disk buffer is $\sim$ 38 TB, corresponding to approximately ten days of continuous data-taking. The resulting RAW files are then copied asynchronously to the off-line storage facilities of the experiment where the reconstruction and analysis of the data are performed.
Fig.~\ref{fig:padmedaq} shows the PADME DAQ system logical scheme.



\subsection{Detector Control System}
The PADME Detector Control System (DCS) is a standalone real-time threaded application running on the DAQ server.
It communicates with the different hardware controllers and sensors of the experiment and continuously monitors all relevant parameters. In case of channel current trips , the DCS can recover it automatically.
To guarantee that the DCS program is running correctly, a dedicated daemon periodically checks its status and, in the case that it has turned off, restarts it.
The list of detector parameters to monitor is passed to the DCS via a configurable text file that specifies the IP address where each parameter can be read asynchronously via Ethernet.
The DCS also monitors the beam and ambient parameters.

The DCS continuously sends all collected information, along with respective timestamps, to the Online Monitor. It also drives the alarm control system and writes all monitored values to the local DCS database.

\subsection{Online Monitor}
Different monitoring tools have been developed to control the detector operation and to check its stability during data-taking.
The parameters to be controlled fall into two categories: ambient parameters, and hardware settings and detector responses.  
The two classes are managed by the different software tools mentioned earlier: the DCS and the DAQ. Both generate data files that are displayed via a web-based Online Monitor.
The basic architecture of this software is client-server and the communication is realized with the REST protocol. 
The graphical characteristics of the plots and the values are interpreted and graphed using the PlotlyJS library. 

This system was used for the entire data collection period of the PADME experiment showing excellent stability and scalability. 
The intrinsic scalability of the system was a key factor in allowing geographically distributed remote shifters to supervise the data acquisition during the 2020 run, when COVID-related travel restrictions prevented collaborators from being able to travel to LNF to assist with data-taking.

\subsection{Offline computing and data storage}

To handle the data transfer from the DAQ system to the final storage area, a Central Data Recording (CDR) software, named PadmeCDR, was developed. 
PadmeCDR runs on a dedicated server connected with a 10 Gbps Ethernet link to the LNF Tier2 infrastructure, part of the Worldwide LHC Computing Grid (WLCG), and automatically transfers the DAQ data to the PADME tape storage area at the tape library of the INFN CNAF national computing center ~\cite{CNAFtape}.

The data are also automatically replicated on the local KLOE2 tape library at LNF as a disaster-recovery safeguard. At the end of the second PADME data taking, a total of $\sim$ 510 TB of data had been written to tape.
The PADME offline computing environment also includes PadmeProd, a custom software package that handles the centralized production of simulated events and the reconstruction of all the data, real and simulated, of the experiment using dedicated WLCG resources located at LNF and CNAF. The distributed CVMFS file system \cite{CVMFS} was used to create a storage area at CNAF from which all WLCG sites can access the PADME simulation and reconstruction software. A Virtual Organization (VO), named {\tt vo.padme.org}, was created to handle the WLCG authorization procedures through the Virtual Organization Management System (VOMS) at CNAF.


\section{PADME offline software}
Data processing and Monte Carlo (MC) simulation use a common software framework. The data format and persistency mechanisms are based on the ROOT ~\cite{ROOT} and \texttt{Geant4}~\cite{GEANT4} libraries. The raw event consists of the full set of digital signals recorded by each channel after zero suppression, along with trigger information (see Sec.~\ref{sec:trg}).

The reconstruction software for each detector converts the raw data into collections of reconstructed hits, after applying decoding procedures based on configuration data to map the electronic channels into the physical elements of the PADME detector. Reconstructed hits contain information about position, time and energy, corrected for alignment and calibration constants (as described in Sec.~\ref{sec:calibration}). During reconstruction, hits are processed with a clusterization logic that creates candidate objects to represent the signature of particles in the detectors. In the ECal and SAC, these are electromagnetic clusters from photons, while in the veto detectors they are clusters of neighboring hits from short tracks of charged particles.

The data from the target are processed differently. 
The summary features of the beam profile are the position of the charge centroid, the width of the distribution in the X and Y view, and the calibrated number of positrons in the bunch, inferred from the total charge released in the detector.

The output format of simulated events consists of collections of \texttt{Geant4} hits for each detector, with position, time, channel identifier and true energy released in the active element of the detector, and the corresponding collections of digitized information. It contains the detector channel hits after some processing aiming at emulating the finite resolution of the detectors and their readout electronics. The reconstruction software converts the digits into collections of reconstructed hits using the same format as for real data, by applying further grouping and smearing in order to best match the detector response. Afterwards, the same clusterization and beam reconstruction algorithms used for real data are also applied to simulated events.

\subsection{Physics processes and detector simulation}
The PADME simulation 
feeds a description of the geometry and materials of the experiment to \texttt{Geant4} for the emulation of all particle interactions. This includes, in addition to the active detectors with their inert support and wrapping materials, the magnet; the vacuum chamber; the beam pipe, starting from a window (made of beryllium in 2019 and Mylar in 2020) that separates the PADME line from the upstream accelerator beam line; and the last bending magnet of the beam before the target. A typical background simulated event consists of approximately 2$\times10^{4}$ beam positrons, with the time structure of a typical beam bunch seen by PADME. The simulation  accounts for the main electromagnetic processes occurring in the interactions of beam positrons in the target at PADME. These processes are bremsstrahlung, two-photon production and Bhabha scattering. Simulating the beam line enables an accurate reproduction of the main features of the beam-induced background. 

For signal acceptance studies, single interactions corresponding to a specific process are generated either by external Monte Carlo generators, in the case of two- and three-photon interactions and other selected electromagnetic processes, or by an internal generator, for example to simulate the production of an invisible dark photon in association with a photon. The final-state particles produced by the generator are injected into the PADME simulation in a production vertex within the target volume and the propagation of one positron of the beam pulse is stopped in the target. The resulting simulation of the signal interaction is fully overlapped with the pile-up from background electromagnetic interactions of the other beam particles.

\subsection{Single- and multi-hit reconstruction}
The front-end signals of all detectors are uniformly sampled inside a readout window, when a trigger signal is received. This window is 1024 ns long for the target and the ECal, and 409.6 ns long for the SAC and the vetoes (see Sec. ~\ref{sec:trg}). The ADC ranges from 0 to 4095 counts with a voltage range of [-0.5, 0.5] V, for bipolar signals, and [0,1] V, for unipolar signals. 

To reduce the amount of processed data, zero-suppression is applied to the vetoes and the ECal, in which waveforms with standard deviation less than 5 mV are not considered for hit reconstruction. In all detectors, a pedestal is calculated on an event-by-event basis from the average of the first few hundred sampled values, before the arrival of the trigger (the exact number changes for each detector), and is then subtracted from each waveform. If a negative polarity signal is obtained, a sign change is applied to the pedestal-subtracted waveform. Finally, the voltage is converted to charge in units of pC. 

The hits are reconstructed from each ADC waveform using different algorithms according to the detector characteristics.
For the BGO crystals in the ECal, overlapping hits in single channels were observed during commissioning, and therefore a multi-hit reconstruction algorithm was developed based on a signal template.
Hit times and energy deposits are determined by evaluating the derivative and the integral of the corresponding shifted and scaled templates, respectively.
To obtain the corresponding energy value a nominal conversion factor of 15.3 pC/MeV (corresponding to 8 photoelectrons per keV and a PMT nominal gain of 1.1$\times$10$^{4}$) is applied. The energy may be corrected for fractional charge-loss due to the signal tail extending beyond the readout window, and for signal saturation. An energy threshold of 1 MeV is applied to the ECal hits.

The SAC crystals and the veto scintillating bars can have multiple hits per event. The time of each hit is therefore determined using a 1D peak finder algorithm based on the TSpectrum class of ROOT with a width of 0.8 ns for the SAC and 2.4 ns for the vetoes. Each peak inside a fiducial time window is then associated with a hit, the time of which is given by the peak position. The SAC hit energy is calculated by integrating the charge over a time interval of 4 ns centered on the peak and converted to MeV with a nominal conversion factor of 0.1 pC/MeV (corresponding to 2 photoelectrons per MeV and a PMT nominal gain of 3.2$\times$10$^{5}$). The veto hit energy is obtained from the peak maximum and converted into MeV with a nominal conversion factor of 25 mV/MeV, using the fact that a minimum-ionizing particle (MIP) crossing a 1 cm scintillator releases approximately 2 MeV of energy.

\subsection{Photon and charged particle reconstruction}
The identification of physics objects, such as photons and charged particles entering a detector, is performed by clustering the hits which are close in time and in space. Appropriate energy thresholds and time differences have been defined for each detector (ECal, SAC, vetos) with MC simulation studies and prototype tests. 
For each detector, the clustering algorithm starts from the list of hits not associated with a previous cluster and with the most energetic hit used as the seed of the new cluster.
For the SAC and the vetoes, hits are not associated with a cluster if there is at least one missing hit between them and the seed along any direction. The energy of the cluster is given by the sum of the energies of all participating hits and the time and the position by the energy-weighted time and position of the hits.
A detailed description of the different clustering algorithms developed for the ECal can be found in Ref.~\cite{Leonardi:2017lvo}.

\section{In-situ detector calibrations}
\label{sec:calibration}
The absolute calibration of detector response and its time alignment is a vital step towards the reconstruction and measurement of physical phenomena.
Being an active element, the target also needs to be calibrated, which is of crucial importance for the determination of the cross section of any physical process.

A good energy calibration is fundamental for the ECal to precisely define the missing mass in dark photon events (of a tenth of MeV), and for the SAC and PVeto to match measured photons with the correct positron to detect bremsstrahlung processes (of few tenths of MeV).
Finally, a good time calibration (of few ns) between the various sub-detectors is also needed to correctly group particles coming from the same interaction.

\subsection{Target calibration}
\begin{figure}[h]
\includegraphics[width=0.5\linewidth]{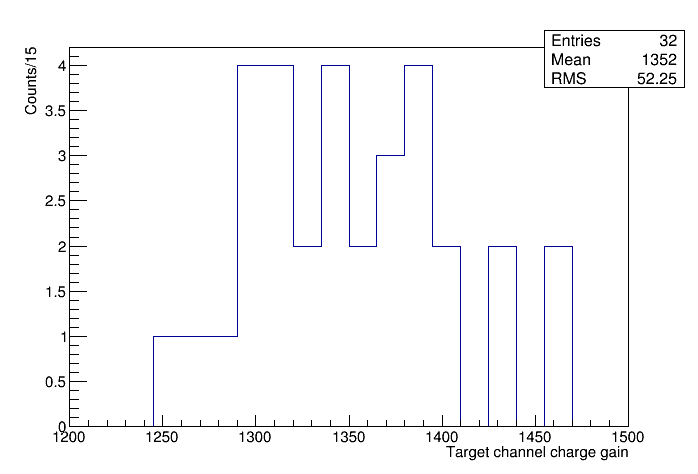}
\includegraphics[width=0.5\linewidth]{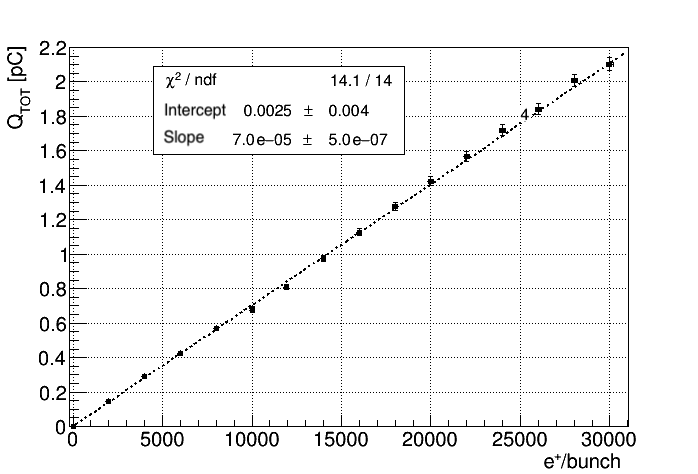}
\caption{
Left: distribution of the charge gains of the target front-end channels measured using the internal charge injection circuit.
Right: active diamond target total charge response to different positron bunch multiplicities.}
\label{fig:target_calibration}
\end{figure}

The diamond target was calibrated by equalizing the response of each amplifier and comparing the amount of charge collected by different strips. 
Each amplifier is equipped with an internal charge injection circuit, which allows the front-end to be tested and the gain to be measured. The charge injected is shared between all the channels selected by an internal shift-register, making
channel-by-channel calibration possible. 
Calibration factors for each channel were inserted in the calibration service of the PADME reconstruction software to estimate the charge collected by the strips (Fig.~\ref{fig:target_calibration}, left). 

The collected charge per bunch is one of the most important pieces of information to extract, because it is expected to be proportional to the bunch multiplicity. An absolute calibration of the diamond target was performed using dedicated runs with different multiplicities of the positron beam as measured by a monitoring lead-glass Cherenkov calorimeter available at the BTF.
At each calibration step, the beam was sent to the BTF calorimeter by turning off the last bending magnet, after which the magnet was turned back on, the beam was sent back to PADME and the target response was recorded. 
Several runs at different bunch multiplicities were taken in order to cross-calibrate the target. In Fig.~\ref{fig:target_calibration}, right, the average of the total charge collected by the X and Y views is shown as a function of the multiplicity measured by the BTF calorimeter. The slope of the linear fit was used to define the absolute calibration factor to be used in the analysis software. 

\subsection{Charged veto calibration}
Gain calibration constants of each channel are derived from data and then used to  perform offline corrections of any differences in the readout hardware of the various detector elements.
In the gain equalization of the charged veto channels is important to apply a uniform threshold and to build clusters of adjacent hits corresponding to MIPs crossing the detector with an angle, and in the case of pile-up to distinguish between multiple charged particles crossing. 

\begin{figure}[h]
\includegraphics[width=0.5\linewidth]{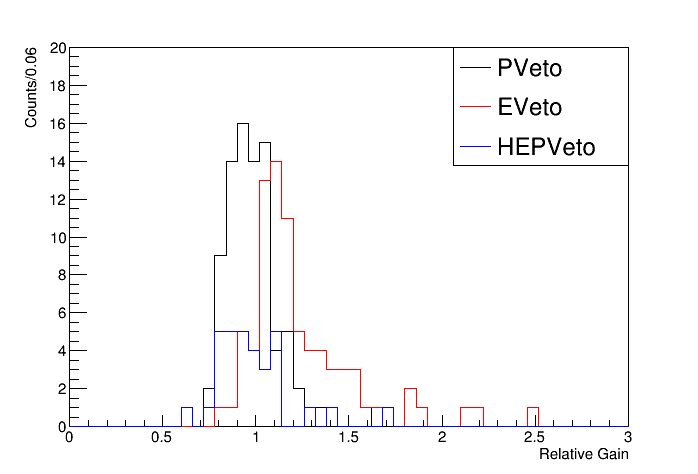}
\includegraphics[width=0.475\linewidth]{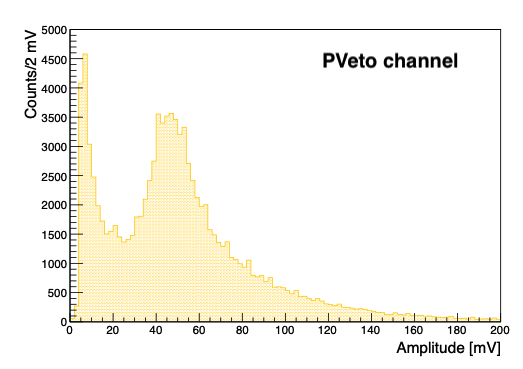}
\caption{
Left: distribution of relative gain of charged veto scintillating bars measured with MIPs.
Right: hit amplitude distribution of a positron veto scintillating bar measured with  positrons in time coincidence with the SAC (bremsstrahlung positrons).}
\label{fig:veto_calibration}
\end{figure}

The relative calibration constants were obtained by evaluating the distribution of the hit energy for each electronic channel of the vetoes, after setting a very low threshold for the hit reconstruction. 
A Gaussian fit around the peak was used to estimate the most probable value of the pulse height distribution averaged with respect to crossing angle and the impact point of the MIP (Fig.~\ref{fig:veto_calibration}, right). The relative calibration constant for each channel was computed by scaling the most probable pulse height to 50 mV (Fig.~\ref{fig:veto_calibration}, left) and applying a scale factor of 34 MeV/V, to obtain a peak energy of 1.7 MeV. This is the expected energy deposit of a MIP crossing 1.1 cm of plastic scintillator.

\subsection{Time calibration}
The time calibration of the front-end channels of downstream detectors (calorimeters and vetoes) is necessary to prevent hits from different 
positron interactions being assigned to the same cluster, which would result in a poor measurement of energy and position.

\begin{figure}[h]
\centering
\includegraphics[width=0.49\linewidth]{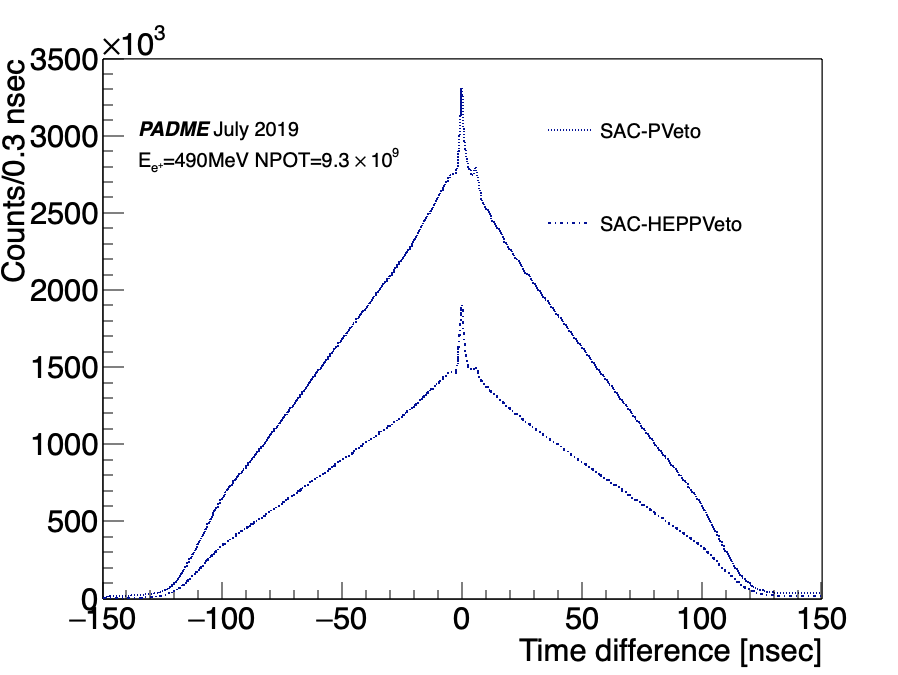}
\includegraphics[width=0.49\linewidth]{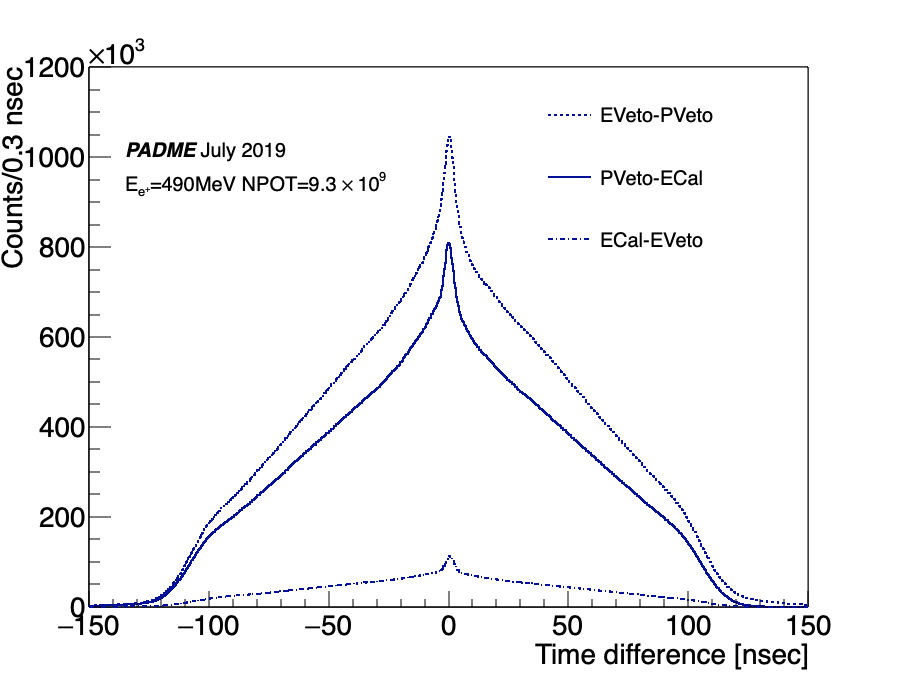}
\caption{Distributions of time differences between different detector channels after time calibration during data-taking with approximately 23000 positrons per bunch.}
\label{fig:timecalibration}
\end{figure}

A strong time correlation exists between adjacent channels of the same detector, because of electromagnetic showers produced by charged particles crossing more than one scintillating bar or crystal. A statistically significant time correlation between channels of different detectors only exists if they are produced by the same primary interaction. The data show clear time correlations between each pair of detectors except the EVeto and the SAC. This is probably due to the fact that the occupancy of the EVeto arises from open-angle primary interactions, such as charged pair-production and showers. Thus, there are few primary interactions expected to produce signals in both the EVeto and the SAC. Because of the lack of correlation between these two detectors, this distribution is not shown in Fig.~\ref{fig:timecalibration}.

The calibration is performed to determine  the time offset t$_{i+1,i}^{D}$ between channel i and the adjacent channel i+1 of detector  D.  Then, the time offset t$_{i+1}^{D}$ with respect to reference channel 0 is calculated by summing the relative calibration constants t$_{k+1,k}^{D}$ of all previous channels k~\textless~i  to align  detector  D in time. 
The final time alignment is obtained by measuring the global time offset for each detector using all hit correlations between PVeto - ECal, EVeto - ECal and SAC - PVeto.
Fig.~\ref{fig:timecalibration}, right, shows the time differences between clusters of different detectors after the calibrations are applied. All distributions have a peak centered at zero (when the physical object comes from the same interaction) above a triangular uncorrelated background (from pile-up events and uncorrelated beam background).

Time differences in detectors have two causes: variation in time-of-flight of particles originating from the same interaction, and different cable lengths or electronics module delays (see Fig.~\ref{fig:timecalibration}). To distinguish the contributions to time differences observed in data, the same procedure was applied to simulated events. This resulted in a second set of time calibration constants reflecting only the difference in time-of-flight of particles, which was subsequently applied to the reconstruction of data.

\subsection{ECal calibration}

Before mounting the calorimeter, all ECal scintillating units (SUs) were calibrated 
with photons emitted by a $^{22}$Na radioactive source~\cite{Albicocco:2020vcy}. This was done to determine the charge vs HV curve, to set the voltage of each unit to produce the same gain. During all the data-taking, the HVs were set to have a charge output of $15.3\,\mbox{pC/MeV}$. This value 
allows signals up to $1\,\mbox{GeV}$ to be fully contained in the digitizer dynamic range. 

\begin{figure}[h]
\centering
\includegraphics[width=0.49\linewidth]{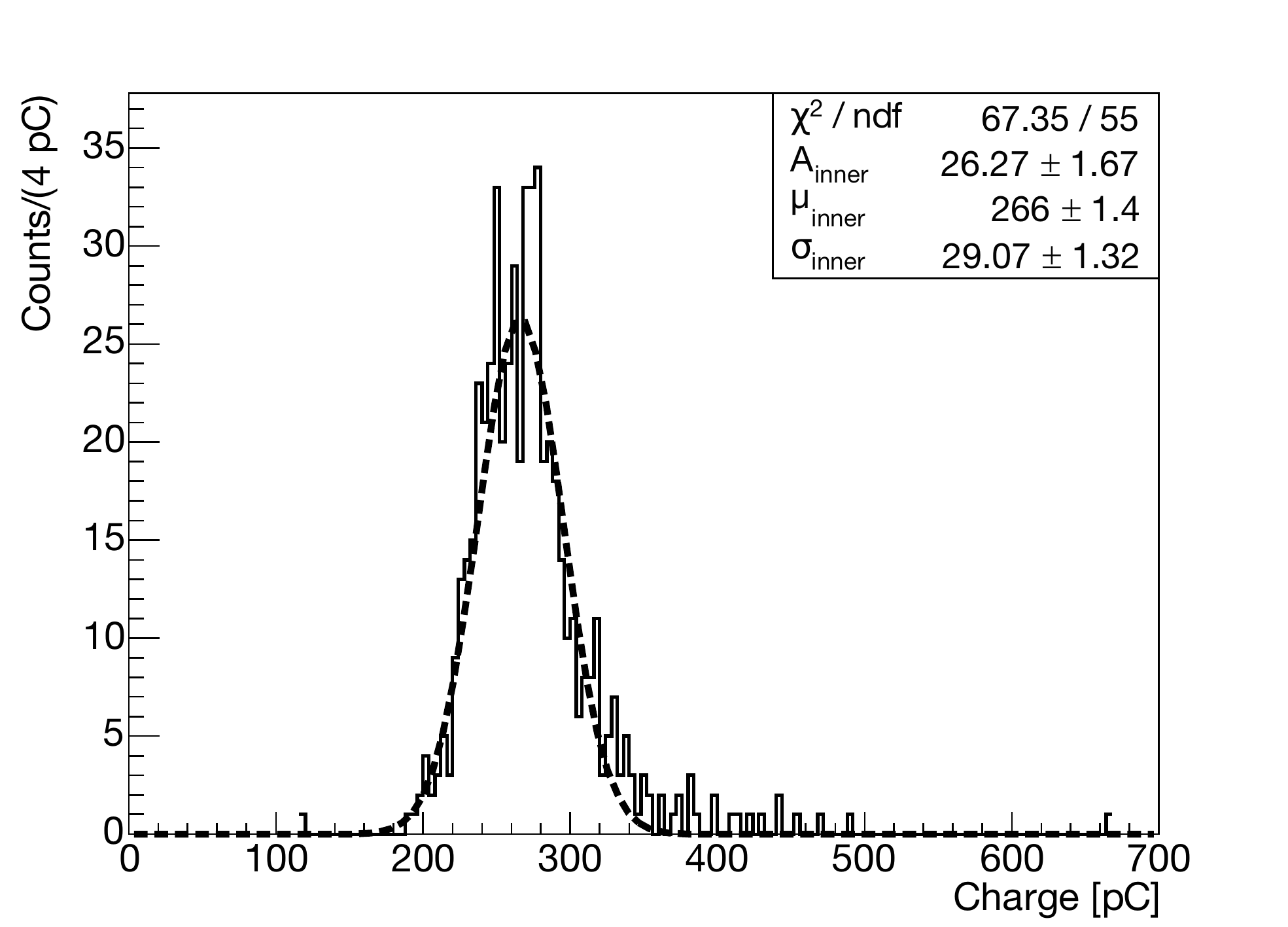}
\includegraphics[width=0.49\linewidth]{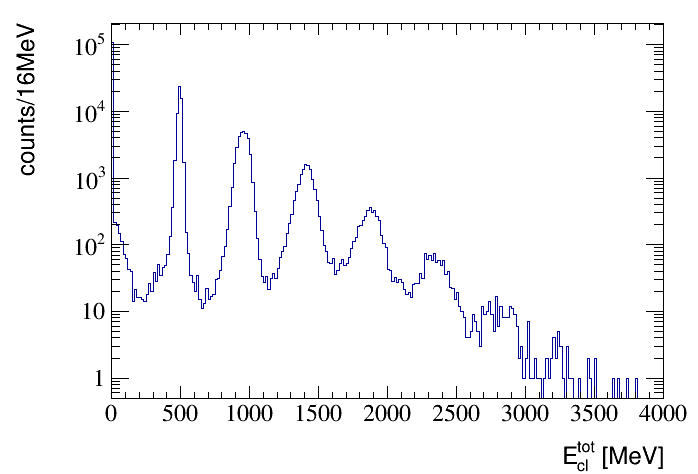}
\caption{Left:  MPVs (Most Probable Values) distribution of all Ecal SUs. The MPVs are obtained fitting each SU charge distribution with Landau curves. Right: distribution of ECal $5\times5$ cluster energies induced by a single-positron beam of $490\,\mbox{MeV}$ energy and reconstructed with the ECal multi-hit algorithm.}
\label{fig:ECal_energycalibration}
\end{figure}

To validate the $^{22}$Na calibration and assess and improve the channel equalisation, cosmic ray events were used. During all the data-taking, the cosmic ray trigger (see Sec.~\ref{sec:trg}) was used to check the calorimeter response to MIPs.  
Fig.~\ref{fig:ECal_energycalibration}, left, shows the distribution of the most probable values from a Landau fit of all SUs charge distributions, resulting from three days of cosmic rays at a rate of few Hz. The superimposed Gaussian fit shows that the ECal response is uniform within ($10.99\pm0.48)\%$.
From the same dataset, the detection efficiencies of the SUs were evaluated: $88.4\%$ of the units showed an efficiency greater than $99.5\%$~\cite{Albicocco:2020vcy}.

Finally, a special run was performed with a single-positron beam of $490\,\mbox{MeV}$ fired directly onto the calorimeter. A $5\times5$ cluster around the crystal with the largest energy deposit was considered and the multi-hit ECal algorithm based on signal template applied.
In Fig.~\ref{fig:ECal_energycalibration}, right, the obtained energy spectrum is shown. The various multiplicity peaks are clearly visible up to seven primary positrons entering the electromagnetic calorimeter, corresponding to a total energy of 3.4 GeV. In normal running conditions the number of hits per crystal per event does not exceed two.  

\subsection{SAC calibration}

The SAC is a Cherenkov-based detector, and therefore its calibration cannot be performed with the same tools used to calibrate ECal crystals. The main difference between the two detectors is in the light output: the light-yield of PbF$_2$ is only $\sim$ 2 photo-electrons/MeV, compared to $\sim$ 200 photo-electrons/MeV for BGO. Hence, the calibration of the SAC was carried out using the BTF positron beam and cosmic rays.

A first set of calibration constants was obtained using beams of single positrons.
By modifying the magnetic fields of the BTF beam line magnets, a 545 MeV single-positron beam was steered to strike the crystals of the SAC. Since the dimensions of the SAC are larger than the ECal hole, only 9 out of 25 crystals were reachable by the beam.

In previous test beams, performed on single PbF$_2$ units before detector assembly, the number of photo-electrons and the photomultiplier gain~\cite{Frankenthal:2018yvf} were found to be $N_{p.e.}$ $\sim$ 2 and $G_i$ $\sim$ 3.2$\times$10$^5$, respectively. 

In data, the collected charge was evaluated with a Gaussian fit of the signals. Combining this information with the parameters found during the test beams, a set of calibration constants was obtained. 

A second set of calibration constants was determined using cosmic rays. 
The energy released in PbF$_2$ by MIPs is 9.373 MeV/cm~\cite{PDG}. Since the short side of a SAC unit is 3 cm long, the energy released by a cosmic ray crossing a crystal perpendicularly is approximately 28.11 MeV. Only vertical cosmic rays crossing all 5 crystals in a column were considered in this analysis. Charge distributions for each crystal were fitted with a Landau curve, and the MPVs from the fit were used to calculate a second set of calibration constants.

\begin{figure}[h]
\includegraphics[width=0.5\linewidth]{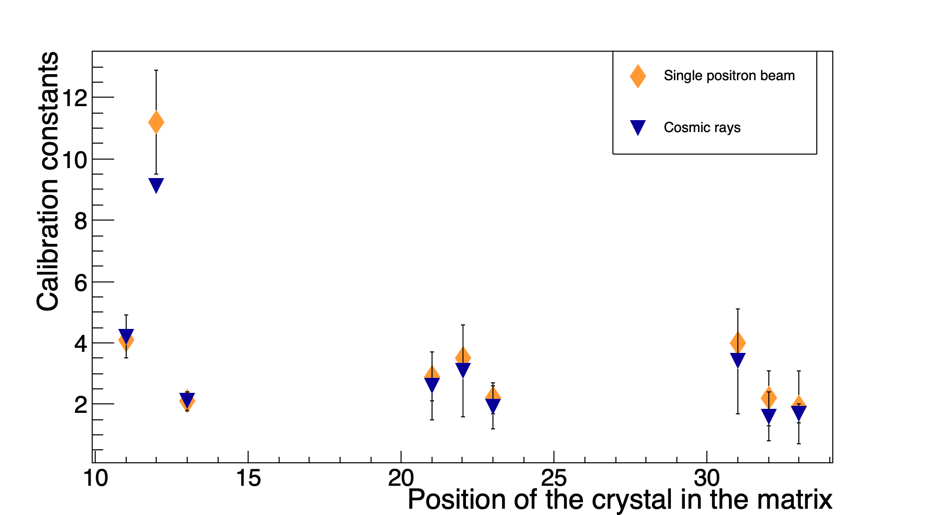}
\includegraphics[width=0.5\linewidth]{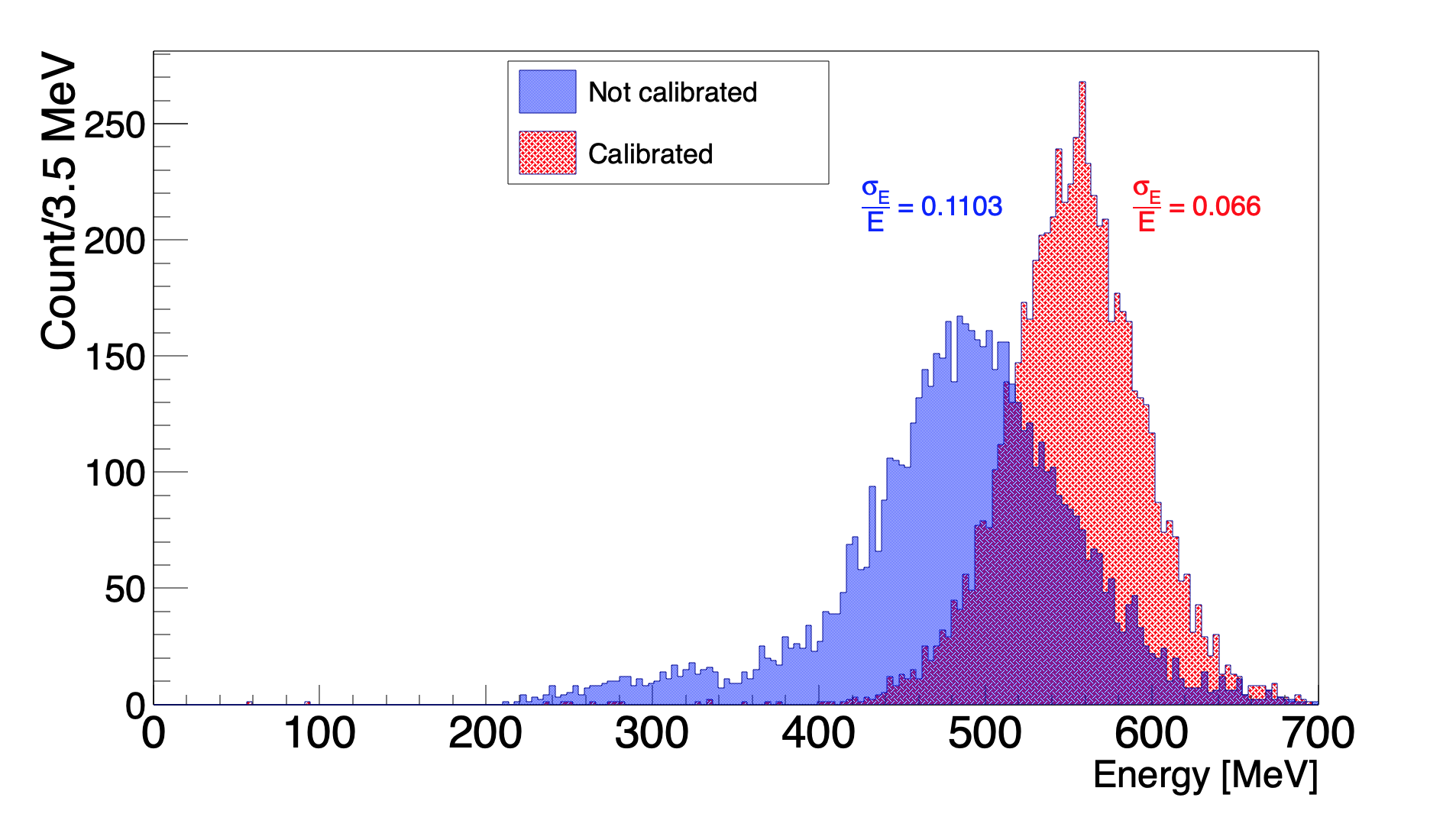}
\caption{Left: SAC calibration constants obtained from single-positron beam measurements (orange) and from cosmic ray measurements (blue).
Right: SAC cluster energy distribution of a 545 MeV single-positron beam, before and after calibration.}
\label{fig:CC_SAC}
\end{figure}

The two sets of calibration constants are plotted in Fig.~\ref{fig:CC_SAC}, left. 
The results of the two independent measurements are compatible within errors. From this work, it was also seen that one of the central crystals (n. 12) was less efficient than the others. The reason was identified in a bad optical contact and was corrected before the data taking of 2020, which is not used for the work presented in this paper.
In Fig.~\ref{fig:CC_SAC}, right, the cluster energy for the 9 central crystals is shown before and after the calibration. 
After calibration, the position of the signal peak is shifted towards the energy of the beam (545 MeV) and the energy resolution improves from 11\% to 6.7\%~\cite{Taruggi:thesis}.

\subsection{Momentum calibration}
\begin{figure}
\centering
\includegraphics[width=0.43\linewidth]{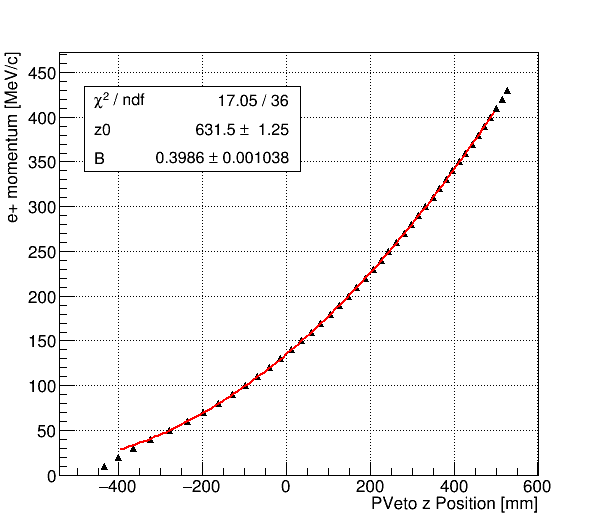}
\includegraphics[width=0.49\linewidth]{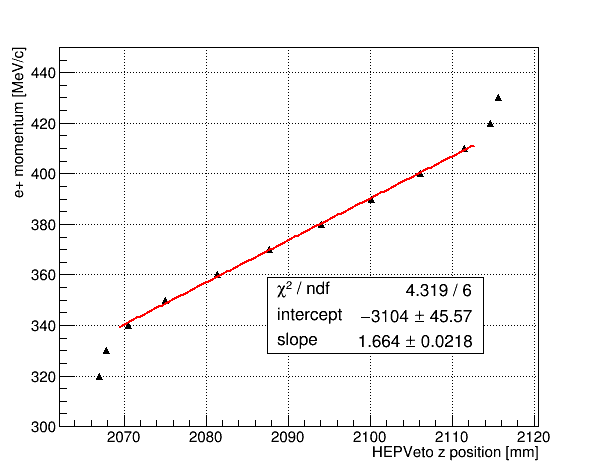}
\caption{Momentum of a positron emerging orthogonally from the target versus impact position along z on PVeto (left) and HEPVeto (right) as obtained from single-positron simulations with a magnetic field of 0.408 T.}
\label{fig:PVETO_momentum_calibration}
\end{figure}

Thanks to the detailed map of the magnetic field,
the veto calibration curve of momentum vs position can be reliably extracted from simulations.
From 50 mono-energetic simulated samples of single positrons, the distribution of cluster z-positions is determined and the momentum calibration curve and resolution are extracted from the average and width of the Gaussian fits. Fig.~\ref{fig:PVETO_momentum_calibration} shows the calibration points obtained from the simulations with the following curve fit superimposed: 
\begin{equation}
p\,(z)=\frac{0.3B\left[{(z+z_{0})}^{2}+x^{2}\right]}{2x},
\end{equation}
where $p$ is the positron momentum, $z$ is the crossing point coordinate on the veto detector and $x$ is the lateral distance of the veto detector from the beam axis (192.5 mm).
The magnetic field has been mapped precisely in 3D by a dedicated survey, and the excitation curve was also measured in a subset of points.
Since the real magnetic field map is used in the simulations, the fit parameters $z_0$ and $B$ of the above simple analytical model, which is exact only for an ideal magnetic field $B$ constant for $z>-z_0$, provide an effective length and a field that reproduce the calibration data points well.

\section{Commissioning with primary positron beam}
The data collected in July 2019 with a primary positron beam represented the completion of the commissioning of the experiment.
Two-photon annihilation processes were clearly visible in the ECal, in addition to the bremsstrahlung processes in the plot correlating PVeto and SAC signals. However, bremsstrahlung events in which the photon goes into the ECal instead of the SAC were only visible in the 2020 data taking with primary beam and the new beam line.

\begin{figure}[h]
\centering
\includegraphics[width=0.49\linewidth]{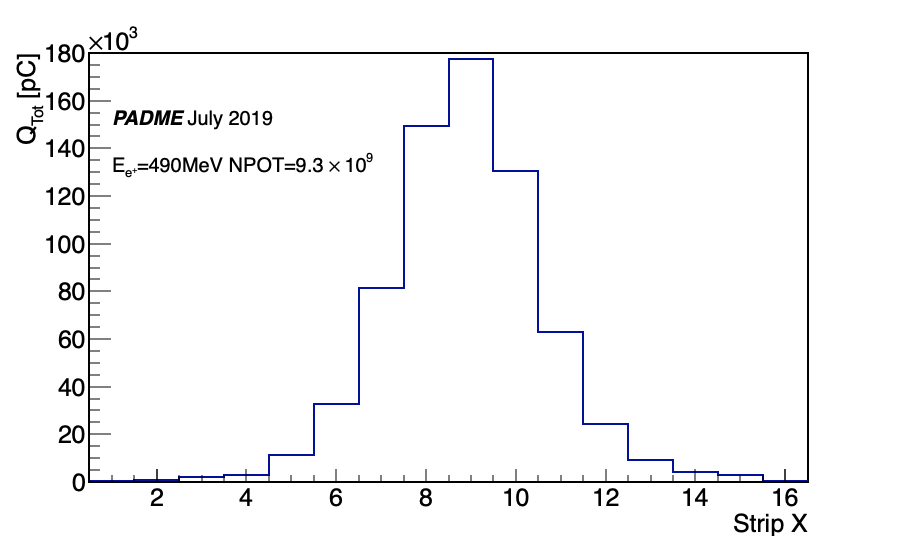} 
\includegraphics[width=0.49\linewidth]{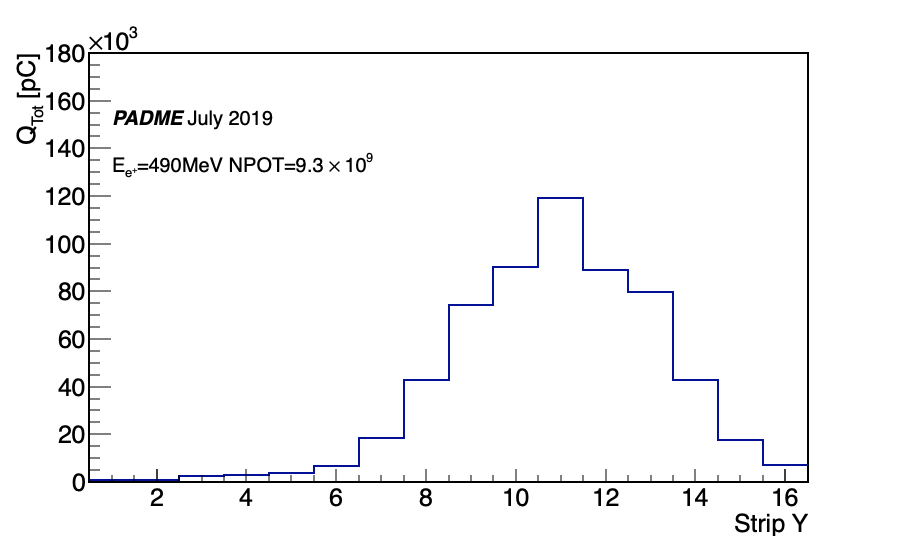} 
\caption{Beam profiles along the X (left) and Y (right) directions as measured by the active diamond target during the July 2019 data-taking. Beam energy and the total number of particles are indicated.}
\label{fig:Target_beambunch}
\end{figure}

In 2019, a primary positron beam of energy 490 MeV and bunch length of 150 ns was used. 
It was focused to a spot of approximately 2-3 mm in X and Y directions on the target, as can be seen in Fig. \ref{fig:Target_beambunch} showing the measured beam profiles. 
The typical positron multiplicity was $\sim$23,000 positrons/bunch (see Fig.~\ref{fig:NPOT_and_NClusters}), corresponding to a density of approximately 150 $e^+$/ns.
This was slightly above the limit imposed by the maximum acceptable event pile-up, which was ascertained to be $\sim$100  $e^+$/ns, after a careful analysis performed during beam commissioning for 2020 data taking.

Thanks to a remote controlled positioning motor, the target can be moved to the ``parked" position where it does not interact with the beam. Runs without target were used to study the beam background arising from beam interactions with different materials (seen by the PADME detectors).
During the commissioning phase the comparison of data with and without target was a powerful tool to investigate the origin of the observed background. The two datasets with and without target consist of 400,000 events each.

\subsection{Detector occupancy}
\begin{figure}[h]
\includegraphics[width=0.5\linewidth]{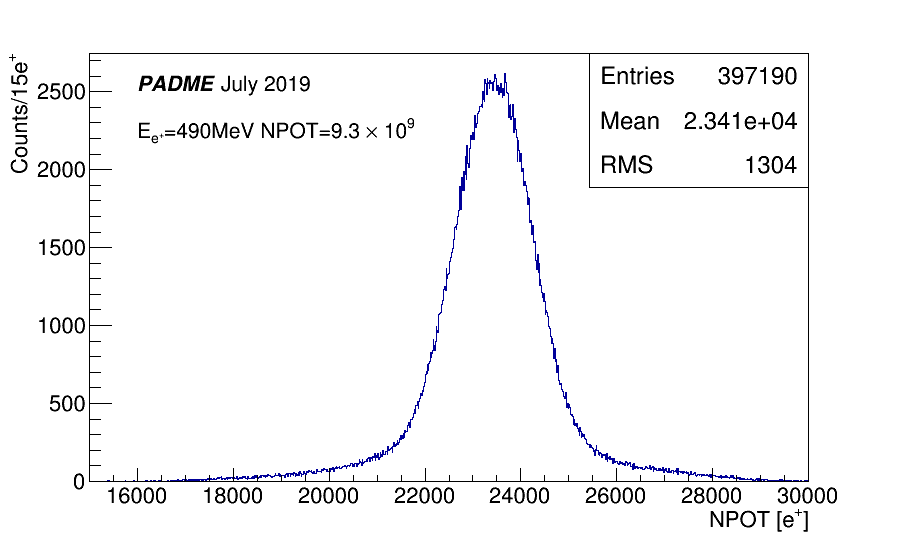}
\includegraphics[width=0.5\linewidth]{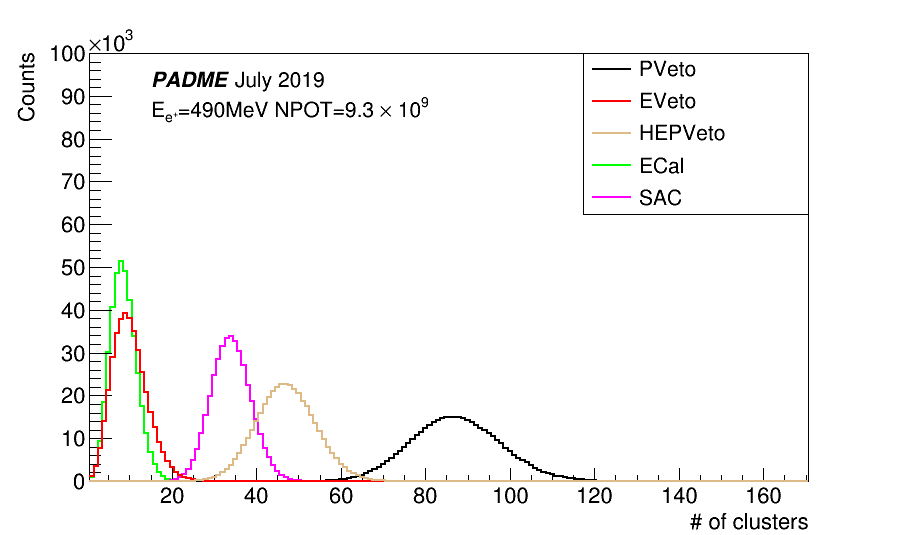}
\caption{Left: distribution of the positron bunch multiplicity as measured by the active diamond target. Right: distribution of the number of clusters for the different detectors. The data correspond to a beam bunch multiplicity of approximately 23,000 positrons per bunch.}
\label{fig:NPOT_and_NClusters}
\end{figure}

With the beam parameters previously mentioned, the distribution of the number of clusters for each detector is shown in Fig.~\ref{fig:NPOT_and_NClusters}.
The number of positrons striking the PVeto is relatively high, while the number of electrons striking the EVeto is very small.
The rate of the photons in the two calorimeters are very different, with ECal registering a much lower rate than SAC, as expected from the hole in the centre of the ECal. The ECal photon flux is dominated by beam background, while the SAC photon flux is mostly due to positron bremsstrahlung in the target.

\begin{figure}[h]
\includegraphics[width=0.5\linewidth]{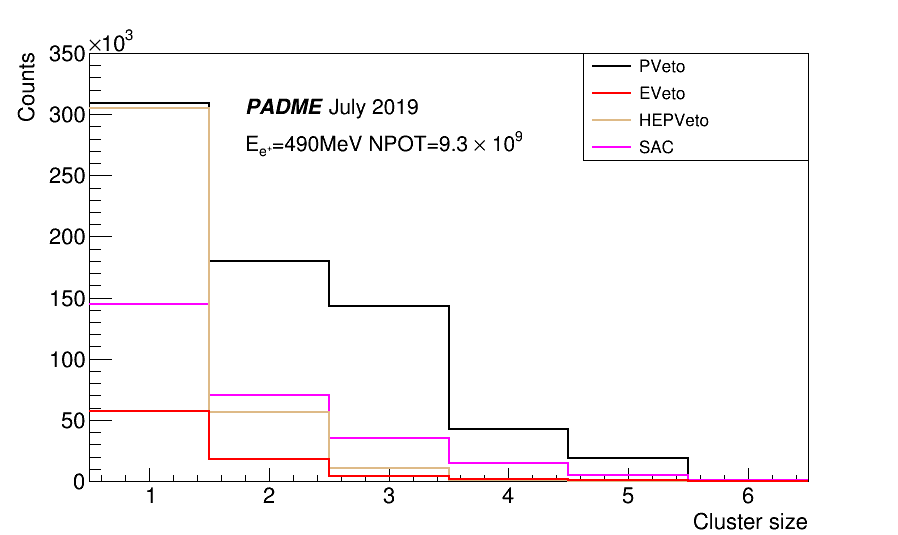}
\includegraphics[width=0.5\linewidth]{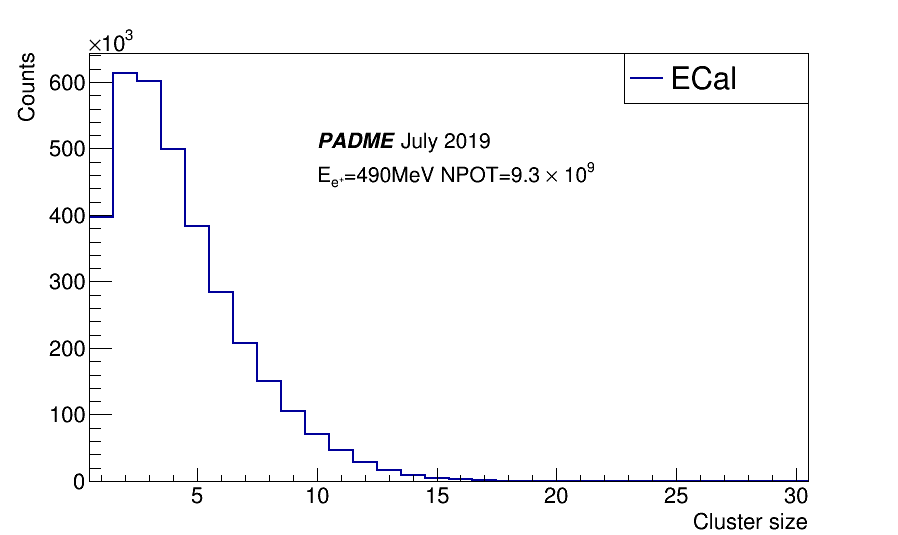}
\caption{Distribution of the cluster size for charged vetos and SAC (left) and ECal (right) during data-taking with approximately 23,000 positrons per bunch.}
\label{fig:ClusterSize}
\end{figure}

To better understand the occupancy of the detectors, in Fig.~\ref{fig:ClusterSize} the distributions of the cluster sizes are shown.
The ECal cluster size is quite large, but a 5x5 matrix is adequate for the clusterization algorithm.
The average cluster size in the PVeto is similar to the EVeto but larger than that of the HEPVeto. This reflects the position and angular orientation
of the experiment, since the HEPVeto can only be struck by positrons that are almost orthogonal to the detector surface. 

\subsection{Photon spectra in the ECal}
\begin{figure}[htbp]
\centering 
\includegraphics[width=.48\textwidth]{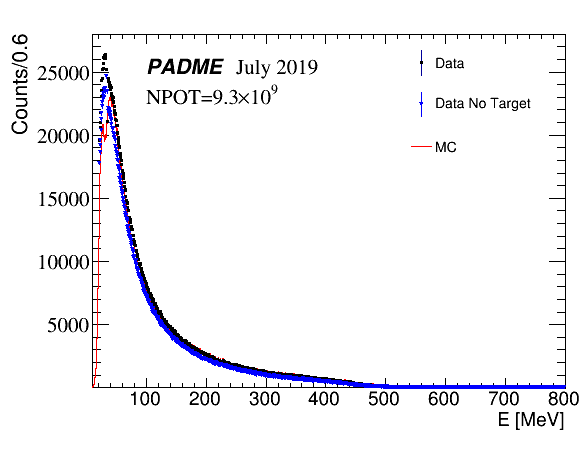}
\includegraphics[width=.48\textwidth]{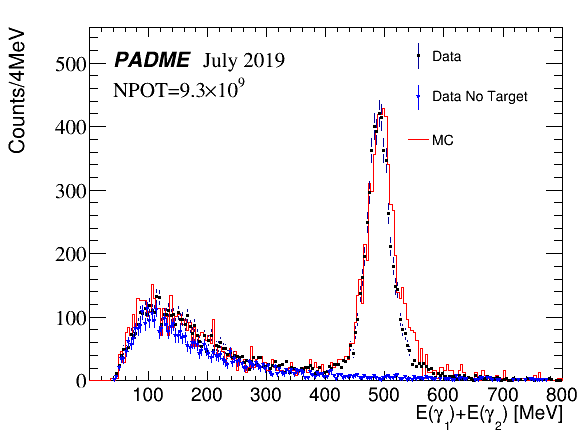}
\caption{\label{fig:EcalCl} 
Energy of the ECal clusters (left) and energy sum of two ECal clusters in time coincidence with clean-up cuts as explained in the text (right). 
The data of a standard run with target (black points) are compared to data of a dedicated run without target (blue points) and to simulation with the 2019 beam line (red line).
The distributions are normalized to the same number of POT equalizing the distribution of out-of-time photons.
The data and simulation correspond to a beam bunch multiplicity of $\sim$23,000 positrons per bunch.
}
\end{figure}

The energy distribution of the clusters reconstructed in the ECal has been studied and compared with simulation (see Fig.~\ref{fig:EcalCl}).
To evaluate the beam-induced background, the spectrum measured in no-target runs is also shown. 
The comparison with the simulation shows good agreement for energies above 30~MeV while the low energy region is dominated by a background component that is also observed in the data collected without target. The source for this background is the large number of low-energy photons produced by unwanted interactions of the beam with beamline and detector materials, a phenomenon that was not predicted during the experimental design.

The most important reference process for PADME is the annihilation of a positron with an electron in the target producing a pair of photons. 
The high rate of bremsstrahlung photons in the forward direction and the coarse granularity of the SAC prevent a clear identification of this process in this detector. Therefore the search for $e^{+}e^{-}\rightarrow\gamma\gamma$ annihilation is performed using only the ECal detector, where the bremsstrahlung background is quite suppressed with respect to the SAC. 
The search exploits the two-photon time coincidence and the conservation of momentum in the transverse plane.
 A simple event selection is implemented by applying the following requirements, without any attempt to optimize the acceptance or the ratio of signal yield to background: 
\begin{itemize}
\item Two ECal clusters in time coincidence $\Delta t =\mid  t_{Cl_1}-t_{Cl_2} \mid < 3 \ \rm{ns}$; 
\item The energy-weighted position of the di-photon pair compatible with the beam direction: $\mid CoG_{x} \mid <1~{\rm cm}$ and $\mid CoG_{y}\mid <1~{\rm cm}$
\footnote{The energy-weighted centroids are defined as
CoG$_{x(y)} = \frac{x(y)_{cl_1}\cdot E_{cl_1}+x(y)_{cl_2} \cdot E_{cl_2}}{E_{cl_1}+E_{cl_2}}$
where  $x(y)_{cl_{1(2)}}$ is the x(y) coordinate of the first (second) ECal cluster and $E_{cl_{1(2)}}$ its energy.};
\end{itemize}
The distribution of the sum of the energies of the clusters that satisfy the above requirements is shown in Fig.~\ref{fig:EcalCl}. The data show a well defined peak at the beam energy, which is indicative of the annihilation process. The data also seem well modeled by the simulation.  
The same distribution obtained for a run without the target is superimposed. As expected in this distribution, the annihilation peak is not present because annihilation can only occur in the target.
Furthermore, the smooth two-photon shoulder matches the shape of this distribution well, indicating that it is almost completely due to beam background coming from interactions of the beam with beam line infrastructure and other materials.

\subsection{Charged particle spectra}
\begin{figure}
\includegraphics[scale=0.27]{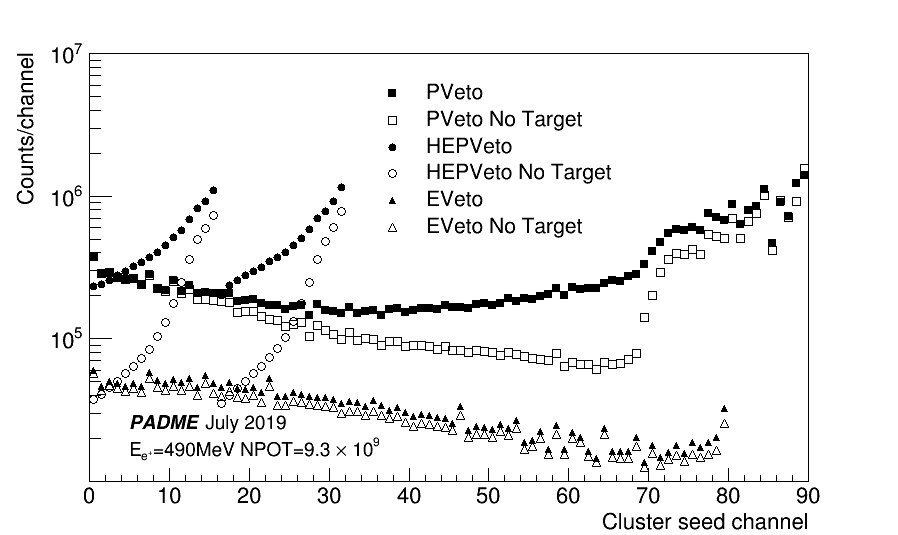}
\includegraphics[scale=0.215]{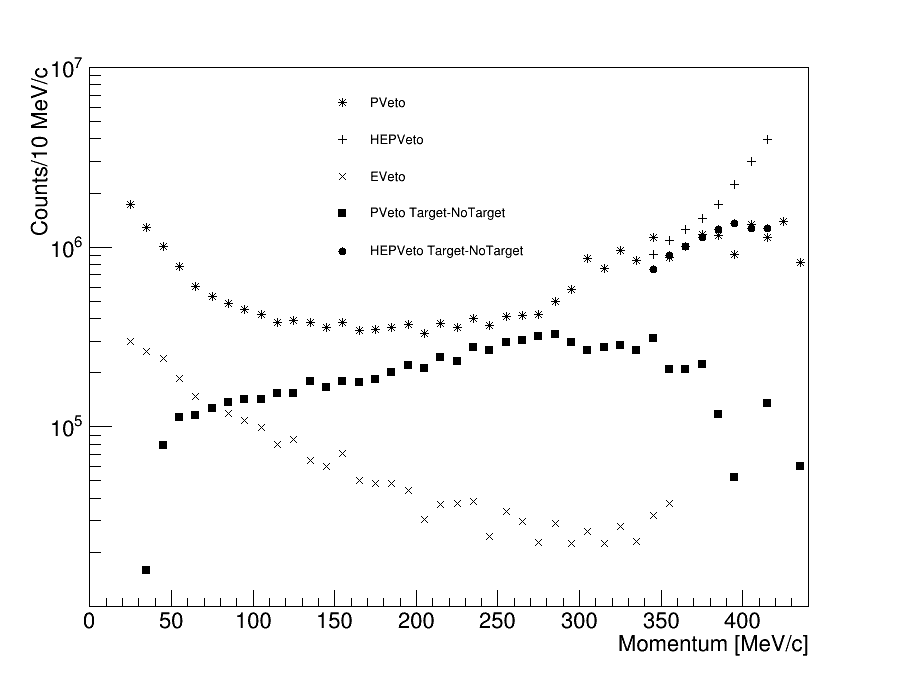}
\caption{Left: charged particle profiles for positron, high energy positron and electron vetos with and without target. Right: corresponding momentum distributions with target data for all vetoes and with no-target data subtracted for PVeto and HEPVeto. The data correspond to a beam multiplicity of $\sim$23,000 positrons per bunch.}
\label{fig:DATA_PositronSpectrum}
\end{figure}

Charged particles coming from the interaction region with energy less than the beam energy are bent towards the three veto detectors.
In Fig.~\ref{fig:DATA_PositronSpectrum} the particle profiles are shown as a function of scintillating bar number for a beam of approximately 23,000 e$^+$/bunch, with and without target.
The normalization of the run without target is obtained by imposing the same occupancy of the first ten scintillator bars of PVeto, where no contributions from interactions in the target are expected. The source of this low-energy charged particle background was removed before 2020 data taking with a more focused beam and larger beam-transport pipes.

For PVeto bars with high channel numbers (those furthest from the target), the rate is very high and the detector is not able to provide precise information.
Here, the high pile-up probability and the large cluster size due to the angle of incidence of the particles on the detector generate many hits in the same bars that
cannot be separated in time.
In this region the hits are dominated by bremsstrahlung interactions in the target and by primary positrons that have slightly less energy than the beam core and that are bent at a larger angle. 
This beam energy spread is mainly caused by the energy degradation of a small fraction of positrons undergoing bremsstrahlung interactions in the beryllium window separating the LINAC and the PADME vacuum. 
This effect was significantly reduced with the new beam line installed in 2020.

The difference in rate with and without the target in the intermediate scintillating bars, approximately between numbers 10 and 70, is due to the contribution of hard bremsstrahlung interactions in the target \cite{PhDOliva}. This corresponds to the momentum range between 50 MeV and 280 MeV, as seen in Fig.~\ref{fig:DATA_PositronSpectrum}.

The HEPVeto detector acceptance partially overlaps that of the last bars of the PVeto.
Thus, the HEPVeto extends the positron veto acceptance by overlapping the region of low efficiency of the PVeto.
The HEPVeto rates with and without target have a different shape for the first scintillating bars, and a more similar shape in the last bars. This provides a measurement of the real positron flux on the PVeto scintillator bars with high  numbers. 
The HEPVeto orientation is almost orthogonal to the positrons, and HEPVeto clusters typically contain relatively few hits, avoiding the saturation occurring in the PVeto.
Consequently, the difference in rates of the HEPVeto with and without target are an estimate of the number of positrons interacting in the target.

As expected, the EVeto rates are much lower than that of the PVeto since electrons must be created either by rather rare primary interactions or in electromagnetic cascades. 
In addition, the EVeto occupancy is similar with and without target, an indication that it is dominated by beam background interactions.

The momentum calibration curves from Fig.~\ref{fig:PVETO_momentum_calibration} can be used to convert the distribution of the charged particles' position along the veto detectors into positron and electron momentum spectra as shown in Fig.~\ref{fig:DATA_PositronSpectrum}. 
Importantly, the momentum is only accurate for forward particles emerging from the target with low transverse momentum.
In the same plot the difference between target and no-target spectra is shown for PVeto and HEPVeto, where it is possible to estimate the rate of positron bremsstrahlung from the target in the range from 50 MeV to 290 MeV, for the PVeto, and from 350 MeV to 390 MeV, for the HEPVeto. This is important since it enables the calibration of the experimental luminosity with another physical process in addition to two-photon annihilation.
The PVeto has a cutoff energy below 50 MeV and excessive occupancy above 290 MeV, while the HEPVeto only begins to saturate above 390 MeV.

\subsection{Photon and charged particle correlation}
\begin{figure}[h]
\includegraphics[width=0.49\linewidth]{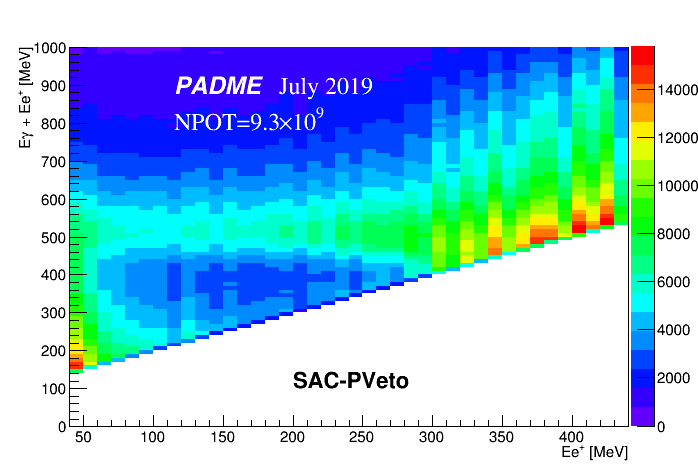}
\includegraphics[width=0.49\linewidth]{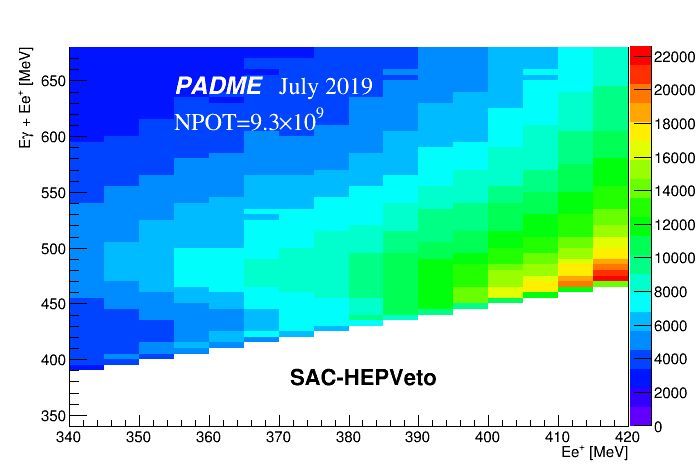}
\caption{Left (right) Scatter plot of the sum of the SAC photon and PVeto (HEPVeto) positron energies against PVeto positron energy evaluated in 1 ns time-coincidence and for SAC photon energy above 100 MeV (50 MeV). The data correspond to a beam multiplicity of $\sim$ 23,000 positrons per bunch.}
\label{fig:SAC_EVETO_momentum_calibration}
\end{figure}

The full detection of positron bremsstrahlung processes occurring in the target ($e^+ Z \rightarrow e^+  \gamma  Z $) is relevant to perform high statistics studies of the correlation between photons and charged particles, which can be used to set the SAC energy scale and to measure photon detection efficiency.
Fig.~\ref{fig:SAC_EVETO_momentum_calibration} shows scatter plots of the sum of photon energy and positron energy with respect to positron energy in a time coincidence window of 1 ns. In the PVeto scatter plot the positron bremsstrahlung signal is clearly visible. This is more difficult to observe in the HEPVeto because of the lower relative energy resolution of the SAC for low energy photons, and to the high beam background flux. 

\section{Conclusions}

PADME is the first fixed-target experiment searching for a dark photon signal by computing the missing mass of single-photon final states resulting from the annihilation of a pulsed positron beam with the electrons of a fixed target. 
The collaboration collected several months' worth of data during 2018 and 2019 with stable secondary and primary positron beams (Run I), which have been analyzed to assess the experimental performance and the beam-induced background level.
Thanks to the detailed analysis performed on this first dataset, a long and stable data-taking period with primary beam and a new beam line was performed in 2020 (Run II).
These Run II data have lower beam related background and they are now being studied to produce the first physics results and 
to establish final detector performances.

In this paper we presented an analysis of the data collected in July 2019 for commissioning the old beam line with a primary beam, and discussed the calibration procedures and the experimental status.
Several detector characteristics have been measured, such as acceptance and energy and time resolutions.
In addition, the simulation software tools, physics object reconstruction algorithms, detector calibration methods and data quality strategy have all been extensively documented.


The active diamond target has operated continuously and stably in vacuum since September 2018. It fulfills its design goals of providing the beam position resolution with better than 1 mm precision for 20,000 positrons/bunch and to measure the integrated luminosity at the level of a few percent after cross-calibration with an external electromagnetic calorimeter.


The electromagnetic calorimeter, which is made of BGO crystals, has excellent energy resolution and an acceptance better than 99.7\% for a 10 MeV mass dark photon signal.
In Run II  a temperature-dependent single-crystal energy calibration was implemented to improve and stabilize the energy resolution for long periods data-taking.


The charged particle scintillating bars attained high efficiency in the detection of MIPs, very good time resolution and good double-hit separation capability. This level of performance was reached thanks to fast front-end electronics, channel gain equalization, and a full multi-hit reconstruction algorithm.


Finally, with a multi-hit reconstruction algorithm for the Small-Angle Calorimeter, the PbF$_{2}$ crystals proved to be efficient for the expected high photon flux. In fact, each crystal was capable of detecting several tens of photons inside a positron bunch of 150 ns length. 

The two-photon annihilation process is clearly visible in  data collected with the primary positron beam in the high resolution electromagnetic calorimeter after loose time and angular selections. 
This is the standard candle for the PADME detector from which it is possible to demonstrate the spatial uniformity of the detector and its stable response over time. 
The bremsstrahlung signal can be seen in both datasets collected with primary and secondary positron beams after accurate time-alignment of the electromagnetic calorimeter and the positively charged particle veto. 



The beam background observed in the data is caused by electromagnetic showers induced by beam halo positrons striking the beam pipe or the finite apertures of the magnet. These particle splashes produce lower energy charged particles and photons with large angular spread in the PADME detectors. Another important contribution to beam background comes from positrons radiating a photon of several MeV while crossing the vacuum-separator beryllium window. 
To handle this background it is necessary to develop efficient multi-hit reconstruction algorithms for all detectors and to perform good time calibrations.


The commissioning process was crucial to prove that the detector design goals have been reached, to optimize the hardware and software tools, and to achieve good agreement between simulation and data both for physics signals, such as bremsstrahlung and two-photon annihilation, and for beam-induced backgrounds. This activity was necessary to address the construction of the new beam line and to establish the position of the new Mylar vacuum separation window for Run II, replacing the beryllim window used in Run I.

\section{Acknowledgments}

We warmly thank the BTF and LINAC teams of LNF for providing an excellent quality 
beam and for their full support during the data-taking period.
This work is partly supported by the Italian Ministry of Foreign Affairs and International Cooperation (MAECI) under the grant PRG00226 (CUP I86D16000060005), the BG-NSF KP-06-DO02/4 from 15.12.2020 as part of MUCCA, CHIST-ERA-19-XAI-009, and TA-LNF as part of STRONG-2020 EU Grant Agreement 824093 projects.

\bibliographystyle{unsrt}
\bibliography{Commissioning.bib}

\end{document}